\documentclass{article}
\usepackage{spconf,amsmath,epsfig,amsfonts}
\usepackage{psfrag}
\usepackage{amsmath}
\usepackage{amssymb}
\usepackage{amsfonts}
\usepackage{color}
\usepackage{tabularx}
\usepackage{subfigure}
\usepackage{pstricks, pst-node, pst-plot, pst-circ}
\usepackage{moredefs}
\usepackage{algorithm,algorithmic}

\newcolumntype{C}[1]{>{\centering\arraybackslash}p{#1}}

\definecolor{MyGrey}{rgb}{0.5,0.5,0.5}

\newcommand{\algorithmicinput}{\textbf{input:}}
\newcommand{\INPUT}{\item[\algorithmicinput]}
\newcommand{\algorithmicoutput}{\textbf{output:}}
\newcommand{\OUTPUT}{\item[\algorithmicoutput]}

\usepackage{subfigure}
\renewcommand{\thesubfigure}{\thefigure.\arabic{subfigure}}
\makeatletter
\renewcommand{\p@subfigure}{}
\renewcommand{\@thesubfigure}{{\bf Fig. \thesubfigure}.\ }
\makeatother

\def\PSNR{\mathrm{ PSNR}}
\def\punit{\, \mathrm}

\title{\vspace{-2mm}Reusing the H.264/AVC Deblocking Filter for Efficient Spatio-Temporal Prediction in Video Coding}
\name{\vspace{-2mm}J\"urgen~Seiler and Andr\'e~Kaup\vspace{-1mm}}
\address{Chair of Multimedia Communications and Signal Processing, \\University of Erlangen-Nuremberg, Cauerstr. 7, 91058 Erlangen, Germany\\
{\{seiler, kaup\}@LNT.de}}

\begin{document}
\topmargin=0mm
%\ninept
\maketitle

%%%%%%%%%%%%%%%%%%%%%%%%%%%%%%%%%%%%%%%%%%%%%%%%%%%%%%%%%%%%%%%%%%%%%%%%%%%%%%%%%%%%%%%%%%%%%%%%%%%%%%%%%

\begin{abstract} \label{abstract}
The prediction step is a very important part of hybrid video codecs for effectively compressing video sequences. While existing video codecs predict either in temporal or in spatial direction only, the compression efficiency can be increased by a combined spatio-temporal prediction. In this paper we propose an algorithm for reusing the H.264/AVC deblocking filter for spatio-temporal prediction. Reusing this highly optimized filter allows for a very low computational complexity of this prediction mode and an average rate reduction of up to 7.2\% can be achieved.
\end{abstract}

%%%%%%%%%%%%%%%%%%%%%%%%%%%%%%%%%%%%%%%%%%%%%%%%%%%%%%%%%%%%%%%%%%%%%%%%%%%%%%%%%%%%%%%%%%%%%%%%%%%%%%%%%

\begin{keywords}
Video coding, Prediction, Deblocking
\end{keywords}

%%%%%%%%%%%%%%%%%%%%%%%%%%%%%%%%%%%%%%%%%%%%%%%%%%%%%%%%%%%%%%%%%%%%%%%%%%%%%%%%%%%%%%%%%%%%%%%%%%%%%%%%%

\section{Introduction} \label{sec:introduction}

Hybrid video codecs like H.264/AVC \cite{Richardson2003} are able to effectively compress video sequences and provide a good viewing experience at manageable data rates. This is achieved by irrelevance reduction on the one hand side and redundancy reduction on the other hand side. Whilst irrelevance reduction results from quantization, the redundancy in a video sequence is removed by prediction, transformations and entropy coding. In this context, the area actually being encoded is always predicted from already transmitted and decoded areas. Afterwards, the prediction residual is transformed and transmitted to the decoder. Since prediction takes place at a very early stage of the encoding, it has a very high influence on the coding efficiency. 

Current video codecs use two strategies for taking advantage of the correlations within a video sequence for prediction. To cope with temporal similarities, motion compensated prediction \cite{Dufaux1995} is used and spatial similarities are exploited by skillfully continuing the signal from already transmitted areas into the area currently encoded. Although the codecs can switch between spatial and temporal prediction for selecting a rate-distortion optimal mode, the fact that a video sequence possesses temporal and spatial correlations at the same time is ignored. Up to now, only few prediction algorithms exist that can make use of spatial as well as temporal redundancies. The algorithms from Jiang \cite{Jiang2009} and Matsuda \cite{Matsuda2010} are two of them to mention.

In \cite{Seiler2008c} we introduced a different spatio-temporal prediction algorithm, the spatially refined motion compensation. This algorithm operates in two stages and uses motion compensated prediction for exploiting temporal correlations and a spatial refinement step for including spatial correlations into the prediction. Even though we already proposed modifications of this algorithms with a reduced complexity \cite{Seiler2010}, the computational load produced by spatial refinement is still far too large for real-time implementations. In order to cope with this, we now propose a method for reusing the H.264/AVC deblocking filter \cite{List2003} for spatio-temporal prediction.

%%%%%%%%%%%%%%%%%%%%%%%%%%%%%%%%%%%%%%%%%%%%%%%%%%%%%%%%%%%%%%%%%%%%%%%%%%%%%%%%%%%%%%%%%%%%%%%%%%%%%%%%%

\section{Spatially Refined Motion Compensated Prediction Principles} \label{sec:prediction}

\begin{figure}
	\psfrag{m}[t][t][0.9]{$m$}%
	\psfrag{n}[t][t][0.9]{$n$}%
	\psfrag{x}[t][t][0.9]{$x$}%
	\psfrag{y}[t][t][0.9]{$y$}%
	\psfrag{t}[t][t][0.9]{$t$}%
	\psfrag{B}[l][l][0.9]{$\mathcal{B}$}%
	\psfrag{R}[l][l][0.9]{$\mathcal{R}$}%
	\psfrag{x0}[t][t][0.9]{$x_0$}%
	\psfrag{y0}[t][t][0.9]{$y_0$}%
	\psfrag{tau}[t][t][0.9]{$\tau$}%
	\psfrag{tau1}[t][t][0.9]{$\tau-1$}%
	\centering
	\includegraphics[width=0.35\textwidth]{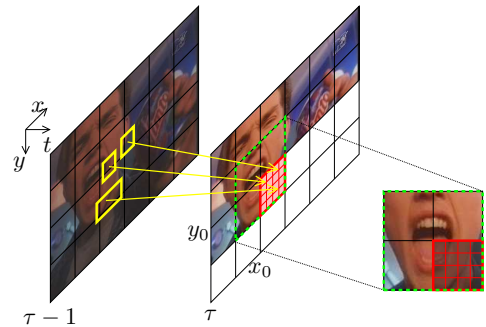}
	\caption{\emph{Relationship between currently predicted macroblock at position $\left(x_0,y_0\right)$ and already decoded neighboring blocks. The predicted block is divided into sub-blocks for motion compensation.}}
	\label{fig:refinement_area}
\end{figure}

Before spatio-temporal prediction with the utilization of the H.264/AVC deblocking filter is introduced, we will briefly review the basic ideas of spatially refined motion compensated prediction. In order to exploit spatial as well as temporal redundancies for prediction, spatially refined motion compensation operates in two stages. During the first stage, pure motion compensated prediction is carried out. Here, all features of motion compensated prediction that have been developed in the recent years like fractional-pel accuracy, multiple reference frames or sub-block partitioning can be used. Using this, the temporal similarities between the individual frames can be exploited well. But, motion compensated prediction takes place without considering spatial redundancies. In order to account for this, the motion compensated signal is spatially refined in the second step. For this, the spatially adjacent, already decoded macroblocks are regarded. If the macroblocks are processed in line-scan order, the macroblocks to the left and above are available. Fig.\ \ref{fig:refinement_area} shows this relationship between the motion compensated block and the already decoded neighboring blocks for the example that the macroblock at position $\left(x_0,y_0\right)$ in frame $t=\tau$ is predicted. In this example, the feature of H.264/AVC to divide a macroblock into sub-blocks for motion compensation is included.

The originally proposed algorithm for spatial refinement \cite{Seiler2008c} obtains the spatio-temporal prediction by iteratively generating a joint model for the motion compensated block and the already decoded neighboring blocks by Frequency Selective Approximation (FSA). The model results from a weighted superposition of two-dimensional basis functions. Although this algorithm yields a significantly increased coding efficiency, it possesses the drawback of a very high computational complexity. In \cite{Seiler2010} an alternative model generation is proposed for an accelerated refinement, but the complexity is still too high for real-time implementations.

In order to cope with this, in the following we introduce an algorithm that reuses a slightly modified version of the H.264/AVC deblocking filter \cite{List2003} for spatial refinement. Due to the reuse of this highly optimized filter, spatial refinement can be carried out at a very high speed.

%%%%%%%%%%%%%%%%%%%%%%%%%%%%%%%%%%%%%%%%%%%%%%%%%%%%%%%%%%%%%%%%%%%%%%%%%%%%%%%%%%%%%%%%%%%%%%%%%%%%%%%%%

\section{Using the H.264/AVC Deblocking Filter for Spatial Refinement} \label{sec:deblocking}

\begin{figure}
\vspace{-0.5cm}
	\begin{center}
		\includegraphics[width=0.35\textwidth]{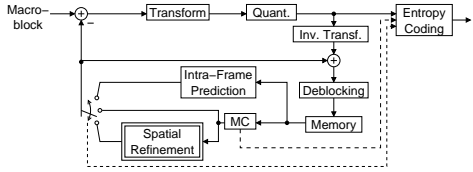}
	\end{center}\vspace{-0.5cm}
	\caption{\emph{Simplified block diagram of a hybrid video encoder with spatial refinement.}}
	\label{fig:coder}
\end{figure}

The original intention of the H.264/AVC deblocking filter \cite{List2003} is to reduce the artifacts which result from quantization. \mbox{Fig.\ \ref{fig:coder}} shows the original position of the deblocking filter in a simplified block diagram of a hybrid video encoder with spatial refinement of the motion compensated predicted signal. The deblocking filter aims at reducing blocking artifacts before the decoded image is copied to the reference buffer from which the motion compensated prediction is carried out. Due to this, using the deblocking filter has three advantages. First, the decoded output image contains less artifacts that might be visible to the viewer. Second, the motion compensated predicted signal fits the input macroblock more, since the motion compensation is performed on the deblocked data. Hence, the coding efficiency is increased. Third, compared to a post-processing deblocking filter, one frame buffer less is required.

Blocking artifacts arise as the individual macroblocks, or even the sub-blocks are quantized independently of each other. Due to this, the pixel values in different blocks differ unnaturally strong and the block boundaries become visible. The deblocking filter aims at detecting unnatural edges at block boundaries and at softening them, if required. 

But, by taking a close look at motion compensated prediction, one can discover that the predicted signal contains similar artifacts. In the process of motion compensated prediction, in the reference frame the best fitting block, or respectively sub-block if macroblock partitioning is used, is determined for the currently regarded block. However, for selecting a block in the reference frame, only the current block is considered and the neighboring blocks are ignored. Hence, the block that is selected for motion compensated prediction may not fit the neighboring, already decoded macroblocks. In this case, the transition between the prediction signal and the neighboring blocks is similar to a blocking artifact. Obviously, such a motion compensated block cannot be a good predictor since the spatial correlations to the neighboring blocks are not considered. For resolving this problem, we propose to use a deblocking filter for spatial refinement of the motion compensated block and for incorporating spatial redundancies into the prediction process.

\begin{figure}
\vspace{-0.5cm}
	\psfrag{p3}[t][t][0.85]{$p_3$}%
	\psfrag{p2}[t][t][0.85]{$p_2$}%
	\psfrag{p1}[t][t][0.85]{$p_1$}%
	\psfrag{p0}[t][t][0.85]{$p_0$}%
	\psfrag{q0}[t][t][0.85]{$q_0$}%
	\psfrag{q1}[t][t][0.85]{$q_1$}%
	\psfrag{q2}[t][t][0.85]{$q_2$}%
	\psfrag{q3}[t][t][0.85]{$q_3$}%
	\psfrag{Sub-block}[t][t][0.8]{Sub-block}%
	\psfrag{boundary}[t][t][0.8]{boundary}%
	\centering
	\includegraphics[width=0.3\textwidth]{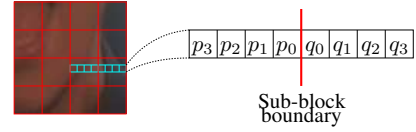}
	\caption{\emph{Samples covered by the deblocking filter.}}
	\label{fig:deblocking}
\end{figure}

The filter used for spatial refinement is similar to the strong deblocking filter from H.264/AVC \cite{List2003} and covers always eight samples, four on the one side of the sub-block boundary and four on the other. As illustrated in Fig.\ \ref{fig:deblocking}, these samples are denoted by $p_0,\ldots, p_3$ and $q_0,\ldots, q_3$. Independently of the actual sub-block partitioning, the filtering is carried out for all sub-block boundaries, first in vertical direction, then in horizontal direction. In this process, the filtering starts with the boundary to the already reconstructed neighboring macroblocks. Using this, spatial similarities can be drawn from the already decoded blocks into the predictor. 

According to \cite{List2003}, two thresholds are required for controlling the amount of filtering. Originally, the thresholds
\begin{eqnarray}
	\alpha &=& 0.8 \left(2^{h/6}-1\right) \\
	\beta &=& 0.5h-7 
\end{eqnarray}
depend on the quantization parameter. Since the amount of deblocking that is required for spatial refinement does not depend on the quantization, the parameter $h$ can be chosen freely. The choice of $h$ will be discussed in Section \ref{sec:results}. Alg.\ \ref{algo:deblocking} lists the steps that are necessary for deriving the filtered values $p^\prime_0,p^\prime_1,p^\prime_2,p^\prime_3,q^\prime_0,q^\prime_1,q^\prime_2,q^\prime_3$ from the original values $p_0,p_1,p_2,p_3,q_0,q_1,q_2,q_3$. Comparing Alg.\ \ref{algo:deblocking} with the original deblocking filter from \cite{List2003} one can discover that the calculations for deriving the filtered values are identical with the only exception that the clipping of the filtered values is not used. As motion compensated samples can differ strongly from the original samples, limiting the range of the filtered values would not be advisable. Due to the similarities, the proposed refinement can run as efficiently as the H.264/AVC deblocking filter and the optimized implementations can be reused. After filtering, the original values of the motion compensated block are replaced by the filtered values. 

\begin{algorithm}[t]
\caption{Deblocking filter for spatial refinement}
\small
\label{algo:deblocking}
\begin{algorithmic}
\INPUT $p_0,p_1,p_2,p_3,q_0,q_1,q_2,q_3$
	\STATE $p^\prime_i=p_i, \ \ q^\prime_i=q_i, \ \ \forall i$
	\IF {$\left|p_0-q_0\right|< \alpha \ \& \ \left|p_0-p_1\right|<\beta \ \& \ \left|q_0-q_1\right|<\beta$}
	\STATE $p_0^\prime = p_0+\left(4\left(q_0-p_0\right)+\left(p_1-q_1\right)+4\right)>\!\!>3$
	\STATE $q_0^\prime = q_0-\left(4\left(q_0-p_0\right)+\left(p_1-q_1\right)+4\right)>\!\!>3$
		\IF {$\left|p_0-p_2\right|<\beta$}
			\STATE $p_1^\prime = p_1+\left(p_2+\left(\left(p_0+q_0+1\right)>\!\!>1\right)-2p_1\right)>\!\!>1$
		\ENDIF
		\IF {$\left|q_0-q_2\right|<\beta$}
			\STATE $q_1^\prime = q_1+\left(q_2+\left(\left(q_0+p_0+1\right)>\!\!>1\right)-2q_1\right)>\!\!>1$
		\ENDIF
	\ENDIF	
	
	\IF {$\left|p_0-q_0\right| < \left(\alpha>\!\!>2\right)+2$}
		\IF {$\left|p_0-p_2\right|<\beta$}
			\STATE $p_0^\prime = \left(p_2+2p_1+2p_0+2q_0+q_1+4\right)>\!\!>3$
			\STATE $p_1^\prime = \left(p_2+p_1+p_0+q_0+2\right)>\!\!>2$
			\STATE $p_2^\prime = \left(2p_3+3p_2+p_1+p_0+q_0+4\right)>\!\!>3$
		\ELSE
			\STATE $p_0^\prime = \left(2p_1+p_0+q_1+2\right)>\!\!>2$
		\ENDIF
		\IF {$\left|q_0-q_2\right|<\beta$}
			\STATE $q_0^\prime = \left(q_2+2q_1+2q_0+2p_0+p_1+4\right)>\!\!>3$
			\STATE $q_1^\prime = \left(q_2+q_1+q_0+p_0+2\right)>\!\!>2$
			\STATE $q_2^\prime = \left(2q_3+3q_2+q_1+q_0+p_0+4\right)>\!\!>3$
		\ELSE
			\STATE $q_0^\prime = \left(2q_1+q_0+p_1+2\right)>\!\!>2$
		\ENDIF
	\ENDIF
	\OUTPUT $p^\prime_0,p^\prime_1,p^\prime_2,p^\prime_3,q^\prime_0,q^\prime_1,q^\prime_2,q^\prime_3$
\end{algorithmic}\vspace{2mm}
\end{algorithm}

These steps are repeated for all vertical and horizontal sub-block boundaries until all motion compensated samples are replaced by the filtered ones. As the already filtered samples are included in the filtering of subsequent samples, the spatial information propagates into the motion compensated block. Finally, the refined block is used for prediction. Since the refined block exploits temporal as well as spatial correlations, it is a better predictor for the block to be encoded and thus increases the coding efficiency. It has to be noted that the proposed usage of the deblocking filter is not intended for replacing the original H.264/AVC deblocking filter. The aim of the original filter is to allay artifacts resulting from quantization whereas the objective of spatial refinement is to exploit spatial as well as temporal redundancies for prediction. As we will show in the next section, the highest coding efficiency can be achieved if spatial refinement and the original deblocking filter are used together.

%%%%%%%%%%%%%%%%%%%%%%%%%%%%%%%%%%%%%%%%%%%%%%%%%%%%%%%%%%%%%%%%%%%%%%%%%%%%%%%%%%%%%%%%%%%%%%%%%%%%%%%%%

\section{Simulations and Results}\label{sec:results} 

\begin{figure}
	\vspace{-3mm}
	\centering
	\input{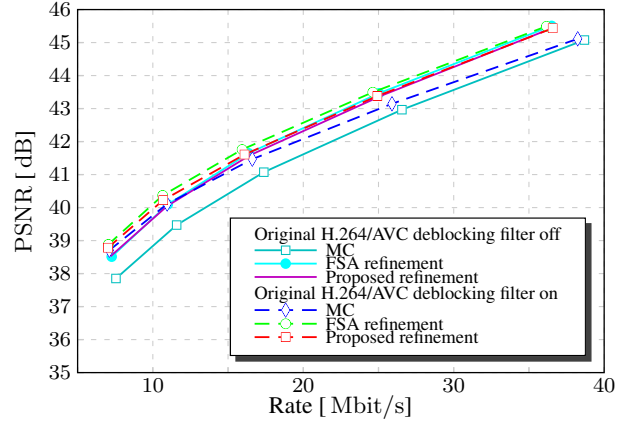}\vspace{-2mm}
	\caption{\emph{Rate-distortion curves for first $399$ P-frames of the 720p-sequence ``Crew''.}}
	\label{fig:rd_curve}
\end{figure}

For evaluating the coding efficiency and the acceleration of the spatial refinement step that can be achieved by the proposed algorithm, we implemented it into H.264/AVC reference encoder JM10.2 running in Main Profile. In addition to the novel refinement, refinement by FSA \cite{Seiler2008c} is considered for comparison. The tests are carried out on four test sequences: two CIF-sequences (``Discovery City'', ``Vimto'') of $100$ frames length and two 720p-sequences (``Crew'', ``Jets'') of $400$ frames length. All sequences are coded in IPPP order and with different quantization parameters (QP). In this process, a wide quality range from QP $16$ to QP $43$ with spacing $3$ is regarded for the CIF-sequences whereas only high qualities with QPs in the range between $16$ and $28$, also with spacing $3$, are regarded for the high resolution sequences. Since not all macroblocks benefit from spatial refinement, one bit per macroblock has to be added as side information to signalize the decoder if spatial refinement should be used or not. Thus, as shown in Fig. \ref{fig:coder}, the encoder now can switch between three modes: spatial prediction, temporal prediction, and spatio-temporal prediction.  We have tested two different scenarios: first, the original H.264/AVC deblocking filter is switched off and the artifacts resulting from quantization are not removed within the prediction loop. Second, the original H.264/AVC deblocking filter is turned on. In this case the deblocking filter is used twice, on the one hand side for reducing quantization artifacts, on the other hand side for spatial refinement of the motion compensated block.

The parameters for spatial refinement by FSA are chosen according to \cite{Seiler2008c}. For the proposed deblocked refinement, only the parameter $h$ can be varied. This parameter controls the amount of filtering, i.\ e.\ at which point a transition is regarded as blocking artifact or as natural edge. For large values of $h$ the thresholds $\alpha$ and $\beta$ also become large and the filtering of the pixels takes place more often than for small values. We have tested different values of $h$ from the range between $10$ and $43$ and discovered that the overall improvement only varies marginally if values from the range between $19$ and $40$ are selected. Hence, we decided to select $h=28$.

Fig.\ \ref{fig:rd_curve} shows the rate-distortion curves of the 720p-se\-quence ``Crew'' for the cases that pure motion compensated prediction is used and that spatial refinement by FSA and the proposed algorithm are included, each time with the original H.264/AVC deblocking filter turned on and off. Taking a look at the pure motion compensated prediction, one can recognize the effect of the original deblocking filter since the coding efficiency is higher for lower qualities. But, with increasing quality the quantization artifacts reduce and the influence of the original deblocking filter becomes void. By regarding the curves for the cases that spatial refinement is used, it becomes apparent that the coding efficiency can be significantly increased by spatial refinement, independently of the actual refinement algorithm. 

In order to condense the results, Table \ref{tab:psnr_gain} lists the average rate reduction or respectively the average $\PSNR$ gain that can be achieved by spatial refinement, compared to pure motion compensated prediction. For this, the Bj{\o}ntegaard metric \cite{Bjontegaard2001} is used. For the case that the original H.264/AVC deblocking filter is turned off, spatial refinement yields very large gains over pure motion compensated prediction. In this case, the blocking artifacts resulting from quantization propagate to the motion compensated prediction, causing a lower coding efficiency. Although spatial refinement can also deal with this problem, it is advisable to switch the original deblocking filter on. In this case, an average rate reduction of up to $15.7\%$ can be achieved by spatial refinement using FSA and up to $7.2\%$ by the proposed deblocked refinement. 

\begin{table}
\fontsize{8}{7}\selectfont
\centering
\begin{tabular}{|l|c|c|c|c|}
\hline
 & \multicolumn{2}{c|}{FSA \cite{Seiler2008c}} & \multicolumn{2}{c|}{Proposed}  \\ \hline
 H.264/AVC&&&&\\ deblocking &  off & on & off & on \\ \hline
 \multicolumn{5}{|l|}{Avg.\ Rate Reduction}\\ \hline
``D.\ City''& $-10.34\%$ & $-5.37\%$ & $-13.65\%$ & $-5.51\%$ \\ \hline
``Vimto'' & $-13.40\%$ & $-5.77\%$ & $-12.15\%$ & $-2.72\%$ \\ \hline
``Crew'' & $-16.89\%$ & $-10.95\%$ & $-15.69\%$ & $-7.22\%$ \\ \hline
``Jets'' & $-22.20\%$ & $-9.97\%$ & $-19.03\%$ & $-6.00\%$ \\ \hline
\multicolumn{5}{|l|}{ Avg.\ $\PSNR$ gain}\\ \hline
``D.\ City'' & $0.74 \punit{dB}$ & $0.35 \punit{dB}$ & $0.99 \punit{dB}$ & $0.36 \punit{dB}$ \\ \hline
``Vimto'' & $0.66 \punit{dB}$ & $0.26 \punit{dB}$ & $0.59 \punit{dB}$ & $0.12 \punit{dB}$ \\ \hline
``Crew''  & $0.81 \punit{dB}$ & $0.45 \punit{dB}$ & $0.75 \punit{dB}$ & $0.29 \punit{dB}$ \\ \hline
``Jets''  & $0.48 \punit{dB}$ & $0.18 \punit{dB}$ & $0.41 \punit{dB}$ & $0.11 \punit{dB}$ \\ \hline
\end{tabular}
\caption{\emph{Average relative rate reduction and average $\punit{PSNR}$ gain, calculated according to \cite{Bjontegaard2001} for spatial refinement by FSA \cite{Seiler2008c} and the proposed algorithm.}}
\label{tab:psnr_gain}
\end{table}

Comparing the results for spatial refinement by FSA and the proposed deblocked refinement, it becomes obvious that FSA always can achieve a considerably higher coding efficiency. But, as mentioned before, FSA is computationally very expensive. Due to all the optimization that has been carried out for developing the H.264/AVC deblocking filter, the proposed refinement algorithm operates very fast. To quantify the speed-up, Table \ref{tab:refinement_time} shows the time per macroblock that is required for spatial refinement. The simulations have been carried out on one core of an Intel Core2 Quad, running at $2.4 \punit{GHz}$ and equipped with $8 \punit{GB}$ RAM. Since the proposed refinement algorithm only requires $0.075 \punit{msec}$ per macroblock it is approximately $900$ times faster than the original FSA refinement and a real-time implementation becomes realistic. This huge acceleration also can justify the lower quality of the spatial refinement

\begin{table}
\fontsize{8}{7}\selectfont
\centering
\begin{tabular}{|l|c|}
\hline
& Processing time\\ \hline
FSA \cite{Seiler2008c} & $67.1 \punit{msec}$\\ \hline
Proposed & $0.075 \punit{msec}$ \\ \hline\hline
Acceleration factor & $895$\\ \hline
\end{tabular}
\caption{\emph{Processing time per macroblock for spatial refinement.}}
\label{tab:refinement_time}
\end{table}

%%%%%%%%%%%%%%%%%%%%%%%%%%%%%%%%%%%%%%%%%%%%%%%%%%%%%%%%%%%%%%%%%%%%%%%%%%%%%%%%%%%%%%%%%%%%%%%%%%%%%%%%%

\section{Conclusion} \label{sec:conclusion} 

Within the scope of this paper we proposed a method for reusing the H.264/AVC deblocking filter for spatio-temporal prediction in video coding. For this, the motion compensated prediction is spatially refined. By exploiting spatial as well as temporal correlations for prediction, a mean rate reduction of up to $7.2\%$ can be achieved by the proposed algorithm. Since the novel algorithm reuses the highly optimized H.264/AVC deblocking filter, the processing time for the refinement is very small and a real-time implementation is possible. 

Although the proposed algorithm already yields an increased coding efficiency, future research has to focus on adapting the deblocking filter to the needs of spatial refinement and to take the actural partitioning of the macroblocks into into account for refinement. 

%%%%%%%%%%%%%%%%%%%%%%%%%%%%%%%%%%%%%%%%%%%%%%%%%%%%%%%%%%%%%%%%%%%%%%%%%%%%%%%%%%%%%%%%%%%%%%%%%%%%%%%%%

%\pagebreak
{%\renewcommand{\baselinestretch}{0.80}
%\footnotesize % changed in spconf.sty

}
\end{document}